\documentclass[aps,prx,superscriptaddress,amsmath,amssymb,floatfix,twocolumn,showpacs]{revtex4-1}
\usepackage{times}
\usepackage{graphicx,graphics,color}
\usepackage[colorlinks,citecolor=blue,linkcolor=red]{hyperref}
\usepackage{color}

\begin{document}

\title{First-principles study of the Kondo physics of a single Pu impurity in a Th host}

\author{Jian-Xin Zhu}
\email[To whom correspondence should be addressed. \\ Electronic address: ]{jxzhu@lanl.gov}
\affiliation{Theoretical Division, Los Alamos National Laboratory,
Los Alamos, New Mexico 87545, USA}
\affiliation{Center for Integrated Nanotechnologies, Los Alamos National Laboratory,
Los Alamos, New Mexico 87545, USA}

\author{R. C. Albers}
\affiliation{Theoretical Division, Los Alamos National Laboratory,
Los Alamos, New Mexico 87545, USA}

\author{K. Haule}
\affiliation{Department of Physics and Astronomy, Rutgers University, Piscataway, New Jersey 08854, USA}

\author{J. M. Wills}
\affiliation{Theoretical Division, Los Alamos National Laboratory,
Los Alamos, New Mexico 87545, USA}

\begin{abstract}

Based on its condensed-matter properties, crystal structure, and metallurgy, which includes a phase diagram with six allotropic phases, plutonium is one of the most complicated pure elements in its solid state.  Its anomalous properties, which are indicative of a very strongly correlated state, are related to its special position in the periodic table, which is at the boundary between the light actinides that have itinerant 5$f$ electrons and the heavy actinides that have localized 5$f$ electrons. As a foundational study to probe the role of local electronic correlations in Pu, we use the local-density approximation together with a continuous-time quantum Monte Carlo simulation to investigate the electronic structure of a single Pu atom that is either substitutionally embedded in the bulk and or adsorbed on the surface of a Th host. This is a simpler case than the solid phases of Pu metal.
For the Pu impurity atom we have found a Kondo resonance peak, which is an important signature of electronic correlations, in the local density of states around the Fermi energy. Furthermore, we show that the peak width of this resonance is narrower for Pu atoms at the surface of Th than for those in the bulk due to a weakened Pu 5$f$-ligand hybridization at the surface.

\end{abstract}
\pacs{74.25.Jb, 74.20.Pq, 71.27.+a, 71.28.+d}

\maketitle

\section{Introduction}
Pure plutonium is one of the most exotic elemental metals with respect to its condensed matter properties, crystal structure, and metallurgy~\cite{OJWick67,DAYoung91,NGCooper00,RCAlbers01,SSHecker04,RCAlbers07}. It has a phase diagram with six allotropic phases, with the low-temperature monoclinic $\alpha$-phase stable up to 395 K, and the technologically important face-center-cubic (fcc) $\delta$-phase stable between 592 Kelvin and 724 Kelvin. While Pu metal exhibits  a significant atomic-volume expansion, with the $\delta$-phase $25\%$ larger in volume than the  $\alpha$-phase, no magnetism has been experimentally observed for the $\delta$-phase volume~\cite{JCLashley:2005}.
These unusual physical and mechanical properties~\cite{NGCooper00} are believed to be caused by the itinerant-to-localized crossover of  5$f$ electronic orbitals~\cite{RCAlbers01}.  Band-structure calculations within the local density approximation (LDA) fail to predict the equilibrium volume of the nonmagnetic $\delta$-phase of Pu~\cite{PSoderlind94,MDJones00}; while the LDA+$U$ calculations gives the experimental $\delta$-phase volume with an appropriate value of $U$ but also indicates an instability toward an antiferromagnetic ground state~\cite{JBouchet00,DLPrice00,SYSavrasov00}. 
Later LDA+$U$ calculations~\cite{AOShorikov05,ABShick05,ABShick06,FCricchio08} that included spin-orbit coupling (SOC) suggested that the magnetism could be naturally quenched and that the correct $\delta$-phase volume could be obtained if the Pu atom is in a closed-shell Pu 5$f^{6}$ configuration,.
However,  the LDA+$U$ method is unable to obtain the major 5$f$-character quasiparticle peak near the Fermi energy as observed by photoemission spectroscopy (PES) on $\delta$-Pu~\cite{AJArko00,LHavela02}. Recently, the LDA+DMFT method~\cite{VIAnisimov97,GKotliar06,KHeld:2007}, which is a combination of LDA with the dynamical mean-field theory (DMFT)~\cite{ThPruschke95,AGeorges96} needed to capture important quantum dynamical fluctuation effects, has become a powerful tool to address the strongly correlated electronic materials. 
When applied to the $\delta$-phase of Pu,  the LDA+DMFT method can explain not only the large volume expansion~\cite{GKotliar06,SYSavrasov01} but can also address the absence of magnetism~\cite{LVPourovskii06,JHShim:2007} and the quasiparticle peak near the Fermi energy~\cite{LVPourovskii:2007,JHShim:2007,AShick07,JXZhu:2007,CAMarianetti:2008,EGorelov:2010}.  The same approach has also recently demonstrated site-selective electronic correlation effects in the $\alpha$-phase of Pu~\cite{JXZhu:2013}. The key to this success lies in the fact that the DMFT generates both the incoherent as well as the coherent parts of the spectral density in the intermediate coupling regime, with the former responsible for atomic-like behavior while the latter for the quenching of magnetism through Kondo screening~\cite{ACHewson:1993}.  

Experimentally, PES has played a significant role in uncovering the correlated-electron phenomena in $f$-electron materials, including elemental Pu solids. This probe measures the occupied part (that is, that below the Fermi energy) of the spectral density. However, to understand the precise nature of the quasiparticle bands in $\delta$-Pu and other Pu-based materials, it would be important to map out the quasiparticle states in the unoccupied regions as well. Scanning tunneling microscopy (STM) has proved to be another important experimental technique for understanding emergent phenomena in strongly correlated electronic systems such as, for example, high-temperature superconductors~\cite{AVBalatsky:2006,OFischer:2007}. This probe measures the local tunneling conductance at the atomic scale. One of the earlier successes of this powerful technique was the imaging of the Kondo resonance state for magnetic adaptors (like Co) on metal surfaces~\cite{VMadhavan:1998,HCManoharan:2000,NKnorr:2002}. 
Recently, this technique has also been successfully extended to  4$f$-based heavy fermion systems~\cite{SErnst:2011,BBZhou:2013} and URu$_2$Si$_2$~\cite{ARSchmidt:2010,PAynajian:2010},  opening up new avenues of approach towards an understanding of $5f$-electron properties in Pu.  No doubt, a clear-cut STM observation of Kondo resonance around a Pu impurity in a nonmagnetic host would provide direct evidence for a correlation-driven entanglement of the Pu 5$f$-electrons with noninteracting conduction electrons in a metallic environment.

While much theoretical work has been done on the electronic correlation effects of the solid phases of Pu, the Kondo physics of a single Pu atom that is either embedded substitutionally in the bulk or adsorbed on the surface of a thorium (Th) host is actually very important as a foundational study for Pu correlation physics, since such a study provides a benchmark for the electronic properties of an isolated Pu impurity in an otherwise non-interacting bath.
In addition, an impurity-like physics plays a critical role within the DMFT methodology, which makes a study of an isolated Pu impurity atom even more valuable. With this in mind, in this article we present calculations of the local electronic structure of Pu impurities in both the bulk and absorbed on the surface. This quantum many-body problem is treated by a combination of the local-density approximation with a continuous-time quantum Monte Carlo method. By considering the competition between the hybridization of the 5$f$ electrons with the delocalized conduction electrons and the local Coulomb repulsion within the Pu-5$f$ electron manifold, Kondo resonances are found in the local Pu-5$f$ spectral density.  By comparing the width of the resonance peak between the geometries, we find the Kondo temperature is higher for Pu atoms substitutionally embedded in the Th bulk than for those adsorbed on the surface of the Th host. Also, due to the strong spin-orbit coupling in the system, the resonance occurs in the $j=5/2$ manifold. 

From the perspective of electronic and magnetic properties, the Pu impurity problem is especially important because of its flexibility to address the electron localization/delocalization of Pu $5f$ electrons. In addition, such an understanding will not only serve as a test of sophisticated theoretical capabilities but will also stimulate local-probe experiments like STM to contribute insights into emergent phenomena in Pu-based materials. 

 \begin{figure}[t]
\centering
\includegraphics[width=0.5\linewidth,clip]{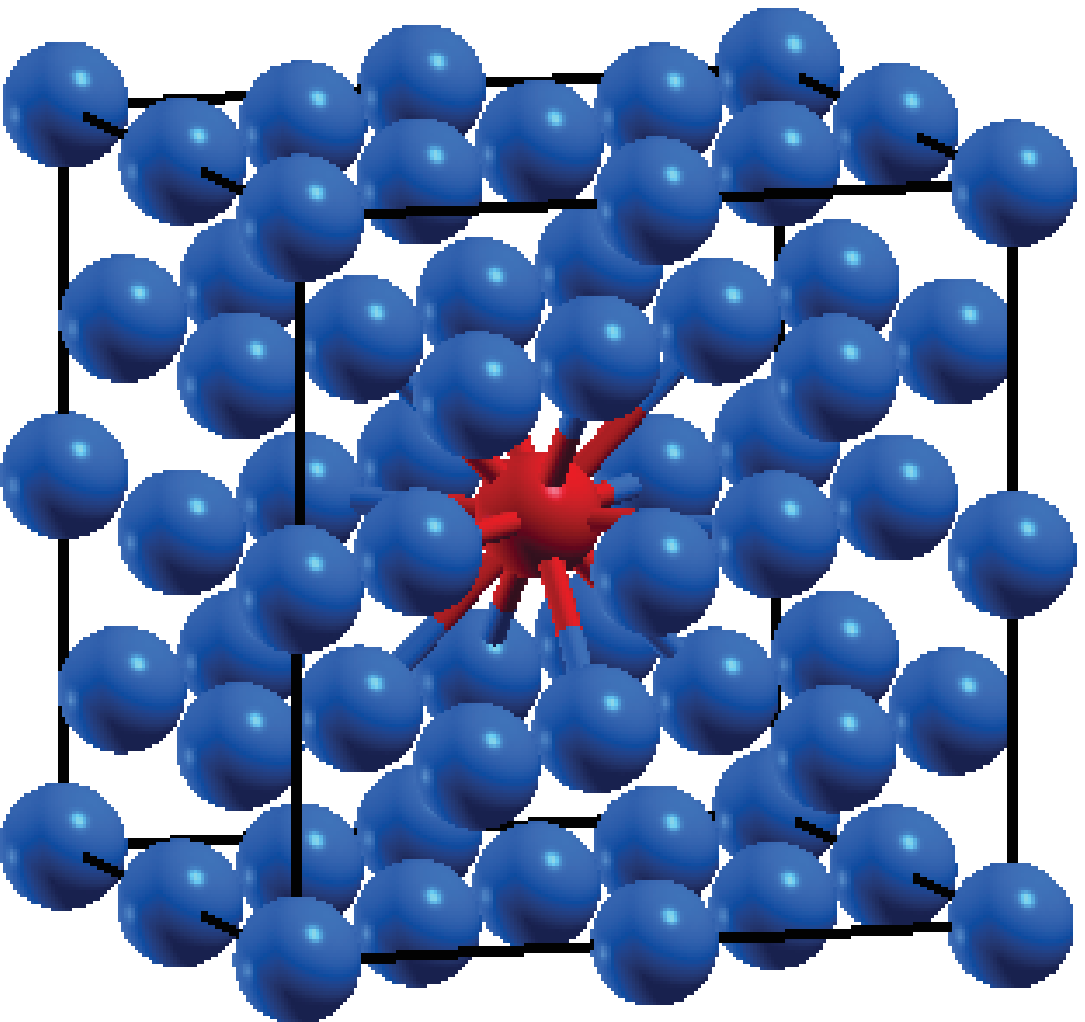}
\includegraphics[width=0.4\linewidth,clip]{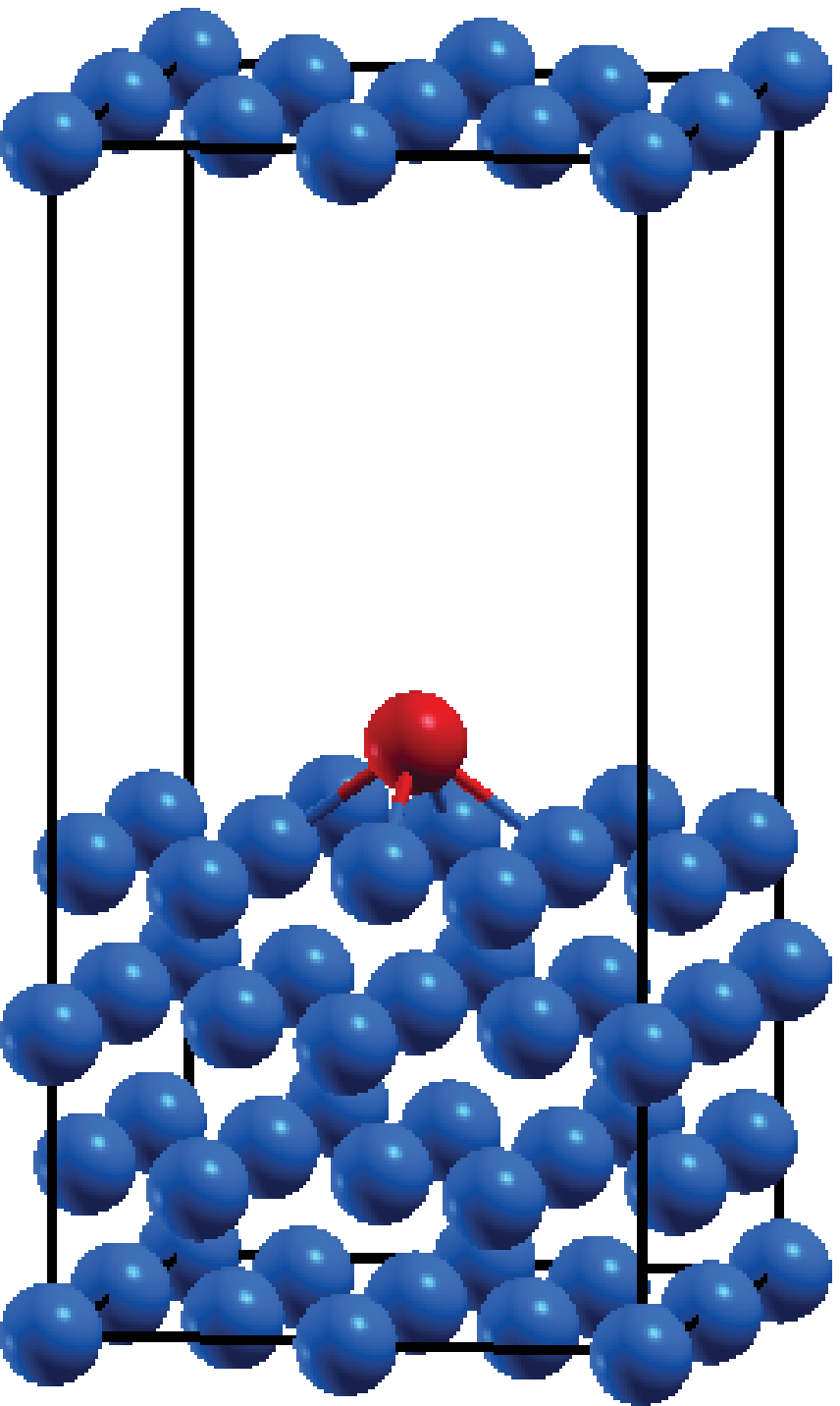}
\caption{(Color online) Geometries of a  Pu impurity substituted in a Th bulk host (left panel) and absorbed on a Th (001) surface (right panel). The Pu atom is marked as a red ball while those of Th are blue balls. The XCrySDen program~\cite{AKokalj:2003}  was used to generate this figure.
}
\label{FIG:struct}
\end{figure}

\section{Theoretical Model and Computational Methods}
A realistic description of the electronic properties of a Pu atom embedded in the bulk or adsorbed on the (001) surface of a Th host, with the geometry shown in Fig.~\ref{FIG:struct}, can be formulated by a multi-orbital Anderson impurity model, with
\begin{equation}
\mathcal{H} = \mathcal{H}_0 + \mathcal{H}_{\text{hyb}} +  \mathcal{H}_{\text{loc}} \;
\end{equation}
and
\begin{subequations}
\begin{eqnarray}
\mathcal{H}_{0}  &=&  \sum_{k} \epsilon_{k} d_{k}^{\dagger} d_{k}\;,\\
\mathcal{H}_{\text{hyb}} & = & \sum_{k,\alpha}(V_{k\alpha} d_{k}^{\dagger} f_{\alpha} + \text{h.c.}) \;,\\
\mathcal{H}_{\text{loc}} & = & \sum_{\alpha} \epsilon_{\alpha} f_{\alpha}^{\dagger} f_{\alpha} + \frac{1}{2} \sum_{\alpha\beta\gamma \delta} U_{\alpha\beta \gamma\delta} f_{\alpha}^{\dagger} f_{\beta}^{\dagger} f_{\delta} f_{\gamma}\;.    
\end{eqnarray}
\end{subequations}
Here, $d_{k}^{\dagger}$ ($d_{k}$) are the creation (annihilation) operators of the conduction-electron continuum for the state $k$ of the Th bath while $f_{\alpha}^{\dagger}$ ($f_{\alpha}$) are those operators of the Pu impurity $f$ electrons with quantum label $\alpha$.  The quantity $\epsilon_{k}$ is the energy dispersion for the Th bath, where the label $k$ includes crystal momentum, band index, and spin, $\epsilon_{\alpha}$ is the on-site single-particle energy levels, where $\alpha$ denotes both the orbital and spin quantum numbers, $U_{\alpha\beta \gamma\delta}$ are the Coulomb matrix elements, and $V_{k\alpha}$ is the strength of hybridization between the local degrees of freedom and the continuum.  

Except for the local Coulomb interaction, our density-functional-theory (DFT) calculations for the Anderson impurity model Hamiltonian (see just above) were done in the generalized gradient approximation (GGA)~\cite{JPPerdew:1996} using the full-potential linearized augmented plane wave (FP-LAPW) method of the WIEN2k code~\cite{PBlaha:2001}. The spin-orbit coupling was included in a second variational
step, with relativistic $p_{1/2}$ local orbitals added to the basis set for the 6$p$ states~\cite{Kunes2001} of plutonium and thorium. The muffin-tin radius was set to 3.3$a_0$ for Pu in the Th bulk and 2.5$a_0$ for Pu on the Th surface, while 2.5$a_0$ was used for all Th atoms in both cases. Here  $a_0$ is the Bohr radius. The  $5\times 5 \times 5$ number of $\mathbf{k}$-points was used for the calculations of Pu in the Th bulk while $7\times 7 \times 3$ for the calculations of Pu on the Th surface. We constructed the supercell by starting with a $2 \times 2 \times 2$ number of  fcc-Th conventional cells with the experimentally determined lattice constant $5.12\;\text{\AA}\;$~\cite{HBohlin:1920}. For the Pu on the Th bulk, Pu was substituted for one Th 
(see the left panel of Fig.~\ref{FIG:struct}) while for the Pu on the Th surface, Pu was optimized along the $z$ direction toward the center of a Th surface plaquette (see the right panel of Fig.~\ref{FIG:struct}). The optimal distance of Pu away from the Th surface is  $1.85\; \text{\AA}\;$  in a vacuum of a $10\; \text{\AA}\;$ perpendicular distance between two neighboring Th slabs.  We note that, because the calculation of systems with a large number of actinide atoms is computationally  demanding in the FP-LAPW method, a full-structure relaxation procedure is not explored. Furthermore, since  the atomic number of Th is close to that of Pu, we expect that a full-structure relaxation will not  qualitatively change the Kondo physics discussed in the present work.  

 To include in the DFT  an explicit of Coulomb interaction between Pu $5f$-electrons  requires a clear definition of the atomic-like local orbitals. In the present work, we use the weight-conserved projection procedure~\cite{KHaule:2010} to extract the local Green's function for the correlated Pu $5f$-orbitals from the full Green's function defined in the DFT basis. Since only the 5$f$ electrons of the Pu impurity site have an onsite Coulomb interaction, we can integrate out the non-interaction bath degrees of freedom from the Th media and arrive at an effective action
\begin{eqnarray}
\mathcal{S}_{\text{eff}} &=& -\sum_{\mu\nu} \int_{0}^{\beta} \int_{0}^{\beta} d\tau d\tau^{\prime} \bar{f}_{\alpha}(\tau) \mathcal{G}_{0,\mu\nu}^{-1}(\tau,\tau^{\prime}) f_{\nu}(\tau^{\prime}) \nonumber \\
&& + \int_{0}^{\beta} d\tau \mathcal{H}_{\text{loc}}[\bar{f}_{\mu}(\tau),f_{\nu}(\tau)]\;.
\end{eqnarray}
Here, the fermion operators in the Hamiltonian become Grassman variables in the action, and $\beta=1/T$ is the inverse of the temperature $T$. The non-interacting impurity Green's function is 
\begin{equation}
\mathcal{G}^{-1}_{0,\mu\nu}(i\omega_n) = (i\omega_n - \epsilon_{\mu}) \delta_{\mu\nu} - \Delta_{\mu\nu}(i\omega_n)\;
\end{equation}
with a hybridization function 
\begin{equation}
\Delta_{\mu\nu}(i\omega_n) = \sum_{k} \frac{V_{k\mu}^{*}V_{k\nu}}{i\omega_n - \epsilon_k}\;,
\label{EQ:hybrid}
\end{equation} 
where we use a Matsubara frequency formalism, with $\omega_n= (2n+1)T\; (n=0,1,\dots)$ for fermions.
A strong-coupling version of the continuous-time quantum Monte Carlo (CT-QMC) method~\cite{PWerner:2006a,PWerner:2006b,KHaule:2007}, which provides numerically exact solutions, was used to solve this multiple-orbital  Anderson impurity problem.  Although the Coulomb interaction parameters can also be determined by DFT calculations, we used a value of $U=4$ eV, which is consistent with previous work on elemental Pu~\cite{SYSavrasov01,JHShim:2007,JXZhu:2007,CHYee:2010}. The remaining Slater integrals, $F^2=6.1$ eV, $F^4=4.1$ eV, and $F^6=3.0$ eV, were calculated using Cowan's atomic-structure code~\cite{RDCowan:1981}, but reduced by 30\% to account for screening. 
 Since the DFT already includes the Hartree-term of the Coulomb interaction, a double-counting correction $E_{dc}=U(n_{f}^{0}-1/2) - J(n_{f}^{0}-1)/2$ with a nominal value of $n_{f}^{0}=5$ for Pu-$5f$ electrons~\cite{CHYee:2010} has been used  in the single-particle $f$-level. For very-well defined local Pu-$5f$ orbitals,  this double-counting scheme has the virtue of numerical stability~\cite{KHaule:2014}.
 The Pu-$5f$ spectral density from the CT-QMC simulations is extracted using the maximal entropy method~\cite{MJarrell96}.
 A similar procedure has been used to study the problem of a transition-metal magnetic atom Co in Cu hosts~\cite{BSurer:2012}, where the formation of a Kondo-resonance Fermi liquid is uncovered. In that kind of systems, since the spin-orbit coupling is negligible,  the cubic harmonics are a good basis to express the self-energy and local Green's function of correlated Co-$3d$ orbitals. In the present work, due to the strong spin-orbit coupling of actinide atoms, we choose the relativistic harmonics $jm_j$ basis to express the self-energy and local Green's function of correlated  Pu-$5f$ orbitals. This choice minimizes the off-diagonal elements of the self-energy and local Green's function, which reduces the minus-sign problem in QMC impurity simulations.

\begin{figure}[t]
\centering
\includegraphics[width=1.0\linewidth,clip]{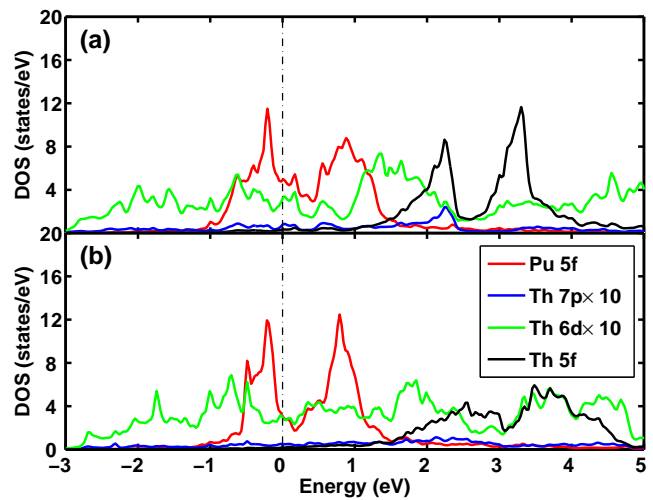}
\caption{(Color online) DFT-based local $5f$ projected density of states (DOS) of the Pu impurity as well as the $7p$, $6d$, and $5f$ projected DOS of the Th sites nearest-neighboring to the Pu impurity  for the cases (a) of Pu in bulk Th and (b) of Pu on the Th surface.  The vertical dashed line in the figure denotes the Fermi energy. 
}
\label{FIG:ldados}
\end{figure}

\section{Results and discussion}
In Fig.~\ref{FIG:ldados}, we show the projected density of states (DOS) on the Pu impurity and that on Th nearest-neighbor sites around the impurity, obtained from DFT-GGA calculations. 
For the Pu atom in bulk Th, the strong spin-orbit coupling of Pu causes the $5f$ states to be split into two manifolds, corresponding to $j=5/2$ and $j=7/2$ of the total angular-momentum quantum number.  These states are located around the Fermi energy in the range from about -1 eV to 0.5 eV for the $j=5/2$ states and from about 0.5 eV to 1.5 eV for the $j=7/2$ states. The Th-$5f$ partial DOS also exhibits SOC-split peak structure, but about 2 eV above the Fermi energy, suggesting that the Th-$5f$ states are not occupied. The Th-$6d$ and Th-$7p$ DOS show wide-band behavior, while the Th-$6d$ contributes dominantly to the bath DOS around the Fermi energy, which hybridizes most strongly with the Pu-$5f$ impurity electronic states. For the Pu impurity on the Th surface, qualitatively similar features are exhibited in the DOS. Noticeably, the Pu-$5f$ DOS shows a narrower quasiparticle peak width with the dominant spectral weight distributed in the range of about -0.6 eV to 0.1 eV for the $j=5/2$ states and 0.5 eV to 1.2 eV for the $j=7/2$ states.  This difference is closely related to the unique structure of  each geometry. For substitutional Pu in bulk Th, the Pu atom has 12 nearest-neighbor Th sites, with a Pu-Th distance of $3.60\;\text{\AA}$; while for Pu atoms on the Th surface, there are only 4 bonds to the nearest Th atoms, with a bond distance of about $3.15\;\text{\AA}$.

\begin{figure}[t]
\centering
\includegraphics[width=1.0\linewidth,clip]{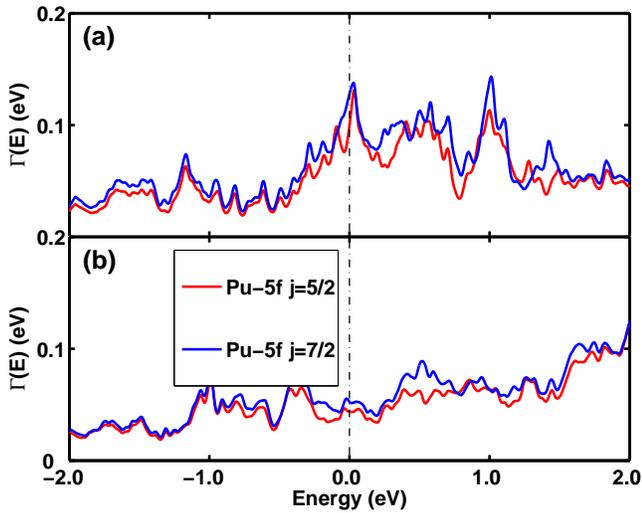}
\caption{(Color online) Hybridization function between the Pu-$5f$ states and the Th bath calculated within the GGA approximation for (a) Pu in bulk Th and (b) for Pu on the Th surface. The vertical dashed line in the figure denotes the Fermi energy. 
}
\label{FIG:ldahybrid}
\end{figure}

To further clarify the close relation between the electronic structure and the local atomic environment for each geometry, we evaluate the frequency-dependent hybridization function, 
$\Gamma(\omega)=-\text{Im}[\Delta(\omega)]/\pi$,  where $\Delta(\omega)$ is the hybridization self-energy given  
by Eq.~(\ref{EQ:hybrid}). We show in Fig.~\ref{FIG:ldahybrid} the hybridization function obtained from the GGA calculations. For the Pu impurity in bulk Th, the hybridization function strength is about 0.8 eV when averaged over  
the frequency range between -0.3 eV and 0.7 eV. For the Pu atom on the Th surface, the strength is about 0.5 eV when averaged over the frequency range between -1 eV and 1 eV. In addition, the overall hybridization function strength 
for both the $j=5/2$ and the $j=7/2$ manifolds is similar, with that for $j=7/2$ being only slightly larger for a given geometry.  We ascribe this slight difference to the fact that the Pu-$5f$ level for 
the $j=7/2$ states, $\epsilon_{j=7/2}$, is closer to the center of gravity of the Th-$6d$ band.

\begin{figure}[b]
\centering
\includegraphics[width=1.0\linewidth,clip]{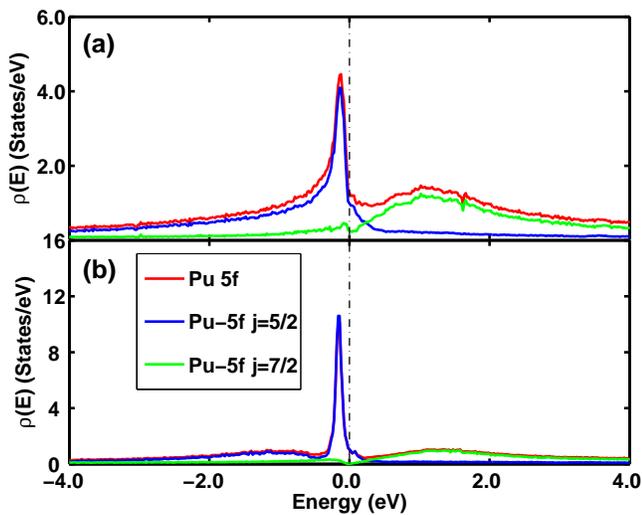}
\caption{(Color online) Total angular-momentum-resolved density of states of Pu impurity in Th (a) and of Pu on the Th surface (b) obtained from the CT-QMC simulations at temperature T = 232 Kelvin.
}
\label{FIG:dmftdos_1}
\end{figure}

By switching on the Coulomb interaction on the Pu impurity, the CT-QMC simulations indicate a significant electronic correlation on the Pu impurity.  For the Pu atom in bulk Th, as shown in Fig.~\ref{FIG:dmftdos_1}(a), the total DOS (red line) obtained from the CT-QMC simulations at $T=232$ Kelvin show not only a quasiparticle peak very close to the Fermi energy but also incoherent peaks at higher energies. From the plot of the $j$-resolved  DOS, one can see clearly that the $j=5/2$ channel (blue line) is the dominant contribution to the quasiparticle peak with only minor intensity from the $j=7/2$ channel at the same energy position (green line), while the incoherent peak located above the Fermi energy appears mainly in the $j=7/2$ channel (green line).  The quasiparticle peak obtained from the CT-QMC simulations is also more than a factor of 2 narrower than the peak on the $j=5/2$ partial DOS obtained from GGA calculations, suggesting a many-body interaction-driven Kondo resonance state. 
For the Pu atom on the Th surface, which has a reduced hybridization, the quasiparticle peak becomes much narrower, as shown by the red and blue lines in Fig.~\ref{FIG:dmftdos_1}(b), especially when compared with the case of a substitutional Pu atom in bulk Th. In addition, the incoherent peak below the Fermi energy also becomes visible in the $j=5/2$ channel.  Therefore, our results indicate that the Kondo resonance state develops in the $j=5/2$ channel.  Physically, it results from almost all of the occupied Pu-$5f$ electrons residing in the $j=5/2$ subshell. When comparing the results from GGA calculations (cf. Fig.~\ref{FIG:ldados}) and from the low-temperature CT-QMC simulations (cf. Fig.~\ref{FIG:dmftdos_1}),  the degree of interaction-driven quasiparticle renormalization between the two geometries is similar.  To quantify the quasiparticle renormalization, we estimate the quasiparticle spectral weight defined as 
\begin{equation}
Z\approx \biggl{[} 1- \left. \frac{\partial \text{Im} \Sigma(i\omega_n)}{\partial \omega_n}\right\vert_{\omega_n=0} \biggr{]}^{-1}\;,
\end{equation}
where the small contribution from the hybridization at the Fermi energy has been neglected. Because the quasiparticle peak occurs in the $j=5/2$ subshell, we consider only the correlation-driven self-energy in this channel, which leads to $Z_{j=5/2}\approx 0.33$ for the Pu in bulk Th while  $Z \approx 0.35$ for the Pu on the Th surface.  When the numerical uncertainty is taken into account,  the above observation is well  supported.

\begin{figure}[t]
\centering
\includegraphics[width=1.0\linewidth,clip]{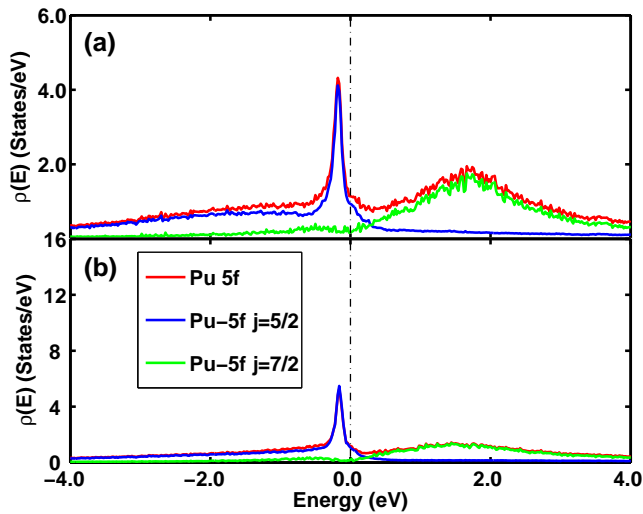}
\caption{(Color online) The same as in Fig.~\ref{FIG:dmftdos_1} except at the temperature T = 725 Kelvin.
}
\label{FIG:dmftdos_2}
\end{figure}

Within the CT-QMC simulations,  the Kondo temperature that is defined from a renormalized fermi-liquid theory for the Anderson impurity model~\cite{ACHewson:2005} that is given by 
\begin{equation}
T_{K} =-\pi Z \text{Im}\Delta(0)/4
\end{equation}
can also be determined.   It was found that $T_{K} \approx 968$ Kelvin for the Pu in bulk Th and $T_{K} \approx 457$ Kelvin for the Pu on the Th surface.  This energy scale should be scaled with the width of the Kondo resonance peak developed near the Fermi energy, which can be measured by STM experiments.  To demonstrate the temperature effect, we also performed CT-QMC simulations at a higher temperature ($T=725$ Kelvin).  As shown in Fig.~\ref{FIG:dmftdos_2}, the Pu-$5f$ spectral density for the case of the Pu in bulk Th is not much different than that at low temperature while the quasiparticle peak  for the case of the Pu on the Th surface is strongly suppressed in intensity.
Interestingly, at this temperature, which is already higher than the Kondo temperature for the Pu on the Th surface, the low-energy DOS peak is still quite visible. One possibility is that the Kondo temperature is systematically underestimated. The other possibility is that the low-energy DOS peak may contain the contribution from both the Kondo-driven many-body physics and the multiplet effect usually occurring in the atomic physics. 
Our  CT-QMC simulations yield a Pu-$5f$ occupancy of $\langle n_f \rangle \approx 5.29$ for the Pu in bulk Th and 5.35 for the Pu on the Th surface. These values suggest that the Pu-$5f$ electronic states are in the mixed-valence regime. The Pu-$5f$ mixed-valence behavior  together with the multiplet effects have also been proposed in the elemental Pu solid~\cite{JHShim:2007,AShick07}. Therefore, this resemblance leads one to believe that the second possibility is more likely. Finally, it is worthwhile to mention that we also performed the CT-QMC simulations for the case of a Pu impurity substituted for a Th atom in the surface of the Th slab. In this geometry, the Pu impurity has 8 nearest-neighboring sites with the Pu-Th distance of $3.60\;\text{\AA}$.  We obtained results (not shown here)  similar  to the case of the Pu on the Th surface.

\section{Summary}
We have performed a first-principles study of the Kondo physics of a Pu impurity in a Th host within the framework of DFT theory in combination with the CT-QMC simulations. By considering two representative geometries of either a substitutional Pu impurity embedded in bulk Th bulk or adsorbed on the Th surface, we have been able to identified the Kondo resonance states developing near the Fermi energy. For a fixed set of on-site Coulomb-interaction parameters, the width of the quasiparticle DOS peak has been shown to be sensitive to the strength of the low-energy hybridization of Pu-$5f$ electrons with the conduction band in the Th medium, as borne out by the two geometries considered here.  By varying the temperature, we have also discussed the nature of the quasiparticle peak in the context of the estimated Kondo temperature scale. With the recent success of STM for exploring emergent phenomena in $f$-electron materials, we hope that the results presented in this work will stimulate STM exploration for Pu-based $f$-electron materials.

\begin{acknowledgments}
This work was supported by U.S. DOE  at
LANL  under Contract No. DE-AC52-06NA25396 through the DOE Office of Basic Energy
Sciences. Part of the theoretical calculations were carried out on a Linux cluster in the Center for Integrated Nanotechnologies, a DOE Office of Basic Energy Sciences user facility.
\end{acknowledgments}


\begin{thebibliography}{99}

\bibitem{OJWick67} O. J. Wick, {\em Plutonium Handbook A Guide to the Technology} (Gordon and Breach, New York, 1967). 

\bibitem{DAYoung91} D. A. Young, {\em Phase Diagrams of the Elements} (University of California Press, Berkeley, 1991).

\bibitem{NGCooper00} {\em Challenges in Plutonium Science}, edited by N. G. Cooper, special issue of Los Alamos Sci.  {\bf 26} (2000).

\bibitem{RCAlbers01} R. C. Albers, Nature {\bf 410}, 759 (2001).

\bibitem{SSHecker04} S. S. Hecker, D. R. Harbur, T. G. Zocco, Porg. Mater. Sci. {\bf 49}, 429 (2004); 
S. S. Hecker, MRS Bull. {\bf 26}, 672 (2001); Metall. Mater. Trans. A {\bf 35}, 2207 (2004).

\bibitem{RCAlbers07} R. C. Albers and Jian-Xin Zhu, Nature {\bf 446}, 504 (2007). 

\bibitem{JCLashley:2005} J. C. Lashley, A. Lawson, R. J. McQueeney, and G. H. Lander, Phys. Rev. B {\bf 72}, 054416 (2005).

\bibitem{PSoderlind94} P. S\"{o}derlind, 
O. Eriksson, B. Johansson, and J. M. Wills, 
Phys. Rev. B {\bf 50}, 7291 (1994).

\bibitem{MDJones00} M. D. Jones, 
J. C. Boettger, R. C. Albers, and D. J. Singh, 
Phys. Rev. B {\bf 61}, 4644 (2000).

\bibitem{JBouchet00} J. Bouchet, 
B. Siberchicot,  F. Jollet, and A. Pasture, 
J. Phys.: Condens. Matter {\bf 12}, 1723 (2000).

\bibitem{DLPrice00} D. L. Price, 
B. R. Cooper, S.-P. Lim, and I. Avgin, 
Phys. Rev. B {\bf 61}, 9867 (2000).

\bibitem{SYSavrasov00} S. Y. Savrasov and G. Kotliar, Phys. Rev. Lett. {\bf 84}, 3670 (2000).

\bibitem{AOShorikov05} A. O. Shorikov, A. V. Lukoyanov, M. A. Korotin, and V. I. Anisimov,  
Phys. Rev. B {\bf 72}, 024458 (2005).

\bibitem{ABShick05} A. B. Shick, V. Drchal, and L. Havela, Europhys. Lett. {\bf 69}, 588 (2005).

\bibitem{ABShick06}  A. B. Shick, L. Havela, J. Kolorenc, V. Drchal, T. Gouder, and P. M. Oppeneer, Phys. Rev. B {\bf 73}, 104415 (2006).

\bibitem{FCricchio08} F. Cricchio, F. Bultmark, and L. Nordstr\"{o}m, Phys. Rev. B {\bf 78}, 100404(R) (2008).

\bibitem{AJArko00} A. J. Arko, J. J. Joyce, L. Morales, J. Wills, J. Lashley, F. Wastin, and J. Rebizant, Phys. Rev. B {\bf 62}, 1773 (2000).

\bibitem{LHavela02} L. Havela, 
T. Gouder, F. Wastin, and J. Rebizant
Phys. Rev. B {\bf 65}, 235118 (2002).

\bibitem{ThPruschke95} Th. Pruschke, M. Jarrell, and J. K. Freericks, Adv. Phys. {\bf 44}, 187 (1995).

\bibitem{AGeorges96} A. Georges, G. Kotliar, W. Krauth, and M. J. Rozenberg, Rev. Mod. Phys. {\bf 68}, 13 (1996).

\bibitem{VIAnisimov97} V. I. Anisimov, A. I. Poteryaev, M. A. Korotin, A. O. Anokhin, 
and G. Kotliar, J. Phys.: Condens. Matter {\bf 9}, 7359 (1997).

\bibitem{GKotliar06} For a review, see G. Kotliar, S. Y. Savrasov, K. Haule, V. S. Oudovenko, O. Parcollet, C. A. Marianetti, Rev. Mod. Phys. {\bf 78}, 865 (2006).

\bibitem{KHeld:2007}  K. Held, Adv. Phys. {\bf 56}, 829 (2007).

\bibitem{SYSavrasov01} S. Y. Savrasov, G. Kotliar, and E. Abrahams, Nature {\bf 410}, 793 (2001).

\bibitem{LVPourovskii06} L. V. Pourovskii, M. I. Katsnelson, A. I. Lichtenstein, L. Hvela, T. Gouder, F. Wastin, A. B. Shick, V. Drchal, and G. H. Lander, Europhys. Lett. {\bf 74}, 479 (2006).

\bibitem{JHShim:2007} J. H. Shim, J. Kaule, and G. Kotliar, Nature (London) {\bf 446}, 513 (2007).

\bibitem{AShick07} A. Shick, J. Kolorenc, L. Havela, V. Drchal, and T. Gouder,
 Europhys. Lett. {\bf 77}, 17003 (2007).
 
\bibitem{LVPourovskii:2007} L. V. Pourovskii, G. Kotliar, M. I. Katsnelson, and A. I. Lichtenstein, Phys. Rev. B {\bf 75}, 235107 (2007). 

\bibitem{JXZhu:2007} Jian-Xin Zhu, A. K. McMahan, M. D. Jones, T. Durakiewicz, J. J. Joyce, J. M. Wills, and R. C. Albers, Phys. Rev. B {\bf 76}, 245118 (2007).

\bibitem{CAMarianetti:2008} C. A. Marianetti, K. Haule, G. Kotliar, and M. J. Fluss, Phys. Rev. Lett. {\bf 101}, 056403 (2008).

\bibitem{EGorelov:2010} E. Gorelov, J.  Kolorenc, T. Wehling, H. Hafermann, A. B. Shick, A. N. Rubtsov, A. Landa, A. K. McMahan, V. I. Anisimov, M. I. Katsnelson, and A. I. Lichtenstein, Phys. Rev. B {\bf 82}, 085117 (2010).

\bibitem{JXZhu:2013}  Jian-Xin Zhu, R. C. Albers, K. Haule, G. Kotliar, and J. M. Wills, Nat. Commun. {\bf 4}, 2644 (2013).

\bibitem{ACHewson:1993} A. C. Hewson, {\em The Kondo Problem to Heavy Fermions} (Cambridge University Press, Cambridge, 1993).

\bibitem{AVBalatsky:2006} A. V. Balatsky, I. Vekhter, and Jian-Xin Zhu, Rev. Mod. Phys. {\bf 78}, 373 (2006).

\bibitem{OFischer:2007} \O. Fischer, M. Kugler, I. Maggio-Aprile, C. Berthod, and C. Renner, Rev. Mod. Phys. {\bf 79}, 353 (2007).

\bibitem{VMadhavan:1998} V. Madhavan, W. Chen, T. Jamneala, M. F. Crommie, and N. S. Wingreen, Science {\bf  280}, 568 (1998).

\bibitem{HCManoharan:2000} H. C. Manoharan, C. P. Lutz, and D. M. Eigler, Nature (Londo) {\bf 403}, 512 (2000).

\bibitem{NKnorr:2002} N. Knorr, M. A. Schneider, L. Diekh\"{o}ner, P. Wahl, and K. Kern, Phys. Rev. Lett. {\bf 88}, 096804 (2002).


\bibitem{SErnst:2011} S. Ernst, S. Kirchner, C. Krellner, C. Geibel, G. Zwicknagl, F. Steglich, and S. Wirth,
Nature (London) {\bf 474}, 362 (2011).

\bibitem{BBZhou:2013} B. B. Zhou, S. Misra, E. H. da Silva Neto, P. Aynajian, R. E. Baumbach, J. D. Thompson, E. D. Bauer, and A. Yazdani, Nat. Phys. {\bf 9}, 474 (2013).

\bibitem{ARSchmidt:2010} A. R. Schmidt,  M. H. Hamidian, P. Wahl, F. Meier, A. V. Balatsky, J. D. Garrett, T. J. Williams, G. M. Luke,
and J. C. Davis, Nature (London) {\bf 465},
570 (2010).

\bibitem{PAynajian:2010}  P. Aynajian, E. H. da Silva Neto, C. V. Parker, Y. Huang, A. Pasupathy, J. Mydosh, and A. Yazdani,
Proc. Natl. Acad. Sci. U.S.A. {\bf 107}, 
10383 (2010).

\bibitem{JPPerdew:1996} J. P. Perdew, K. Burke, and M. Ernzerhof,
Phys. Rev. Lett. {\bf 77}, 3865 (1996).

\bibitem{PBlaha:2001} P. Blaha, K. Schwarz, G. K. H. Madsen, D. Kvasnicka, and J. Luitz
{\em WIEN2k: An Augmented Plane Wave + Local Orbitals Program for Calculating Crystal Properties} (K. Schwarz, Tech. Universit\"{a}t Wien, Austria, 2001).

\bibitem{Kunes2001} J. Kune{\u s}, P. Nov{\'a}k,  R. Schmid, P. Blaha, and K. Schwarz,
Phys. Rev. B {\bf 64}, 153102 (2001).

\bibitem{HBohlin:1920} H. Bohlin, Ann. Phys. (Leipzig)  {\bf 61}, 421 (1920).

\bibitem{KHaule:2010} K. Haule, C.-H. Yee, and K. Kim, Phys. Rev. B {\bf 81}, 195107 (2010).


\bibitem{AKokalj:2003} A. Kokalj, Comput. Mater. Sci. {\bf 28}, 155 (2003).

\bibitem{PWerner:2006a} P. Werner,  A.  Comanac, L.  de Medici, M.  Troyer,  and A. J.   Millis, 
Phys. Rev. Lett. {\bf 97}, 076405 (2006).

\bibitem{PWerner:2006b}  P. Werner and  A. J. Millis, 
Phys. Rev. B {\bf 74}, 155107 (2006).

\bibitem{KHaule:2007}  K. Haule, 
Phys. Rev. B {\bf 75}, 155113 (2007).

\bibitem{CHYee:2010} C.-H. Yee, G. Kotliar, and K. Haule, Phys. Rev. B {\bf 81}, 035105 (2010).

\bibitem{RDCowan:1981} R. D. Cowan, {\em The Theory of Atomic Structure and Spectra} (University of California Press, Berkeley, 1981).

\bibitem{KHaule:2014} K. Haule, T. Birol, and G. Kotliar, Phys. Rev. B {\bf 90}, 075136 (2014).

\bibitem{MJarrell96} M. Jarrell and J. E. Gubernatis, Phys. Rep. {\bf 269}, 133 (1996).

\bibitem{BSurer:2012} B. Surer, M. Troyer, P. Werner, T. O. Wehling, A. M. L\"{a}uchli, A. Wilhelm, and A. I. Lichtenstein, Phys. Rev. B {\bf 85}, 085114 (2012).

\bibitem{ACHewson:2005} A. C. Hewson, J. Phys. Soc. Jpn. {\bf 74}, 8 (2005).

\end{thebibliography}
\end{document}